\documentclass[twocolumn]{aastex631}

\usepackage{lineno}
\usepackage{makecell}

\newcommand{\hotpants}{{\tt Hotpants}}

\begin{document}

\title{Optimizing Convolution Direction and Template Selection for Difference Image Analysis}

\author[0000-0001-5281-731X]{Rodrigo Angulo}
\affiliation{Johns Hopkins University, William H. Miller III Department of Physics and Astronomy, 3400 North Charles St., Baltimore, MD 21218, USA}

\author[0000-0002-4410-5387]{Armin Rest}
\affiliation{Johns Hopkins University, William H. Miller III Department of Physics and Astronomy, 3400 North Charles St., Baltimore, MD 21218, USA}
\affiliation{Space Telescope Science Institute, 3700 San Martin Dr., Baltimore, MD 21218, USA}

\author[0000-0003-2379-6518]{William P. Blair}
\affiliation{Johns Hopkins University, William H. Miller III Department of Physics and Astronomy, 3400 North Charles St., Baltimore, MD 21218, USA}

\author[0000-0001-5754-4007]{Jacob Jencson}
\affiliation{IPAC, Mail Code 100-22, Caltech, 1200 E.\ California Blvd., Pasadena, CA 91125}

\author[0000-0003-4263-2228]{David A. Coulter}
\affiliation{Space Telescope Science Institute, 3700 San Martin Dr., Baltimore, MD 21218, USA}

\author[0000-0001-5233-6989]{Qinan Wang}
\affiliation{Johns Hopkins University, William H. Miller III Department of Physics and Astronomy, 3400 North Charles St., Baltimore, MD 21218, USA}

\author[0000-0002-2445-5275]{Ryan~J.~Foley}
\affiliation{University of California, Santa Cruz, 1156 High St., Santa Cruz, CA 95064}

\author[0000-0002-5740-7747]{Charles~D.~Kilpatrick}
\affiliation{Center for Interdisciplinary Exploration and Research in Astrophysics (CIERA) and Department of Physics and Astronomy, Northwestern University, Evanston, IL 60208, USA}

\author[0000-0002-0514-5650]{Xiaolong Li}
\affiliation{Johns Hopkins University, William H. Miller III Department of Physics and Astronomy, 3400 North Charles St., Baltimore, MD 21218, USA}
\affiliation{LSST-DA Catalyst Fellow}

\author[0000-0002-7559-315X]{C\'esar Rojas-Bravo}
\affiliation{University of California, Santa Cruz, 1156 High St., Santa Cruz, CA 95064}
\affiliation{School of Astronomy and Space Science, University of Chinese Academy of Sciences, Beijing 100049, People’s Republic of China}
\affiliation{National Astronomical Observatories, Chinese Academy of Sciences, Beijing 100101, People’s Republic of China}

\author[0000-0001-6806-0673]{Anthony L. Piro}
\affiliation{The Observatories of the Carnegie Institution for Science, 813 Santa Barbara Street, Pasadena, CA 91101}


\begin{abstract}
Difference image analysis (DIA) is a powerful tool for studying time-variable phenomena, and has been used by many time-domain surveys. Most DIA algorithms involve matching the spatially-varying PSF shape between science and template images, and then convolving that shape in one image to match the other. The wrong choice of which image to convolve can introduce one of the largest sources of artifacts in the final difference image. We introduce a quantitative metric to determine the optimal convolution direction that depends not only on the sharpness of the images measured by their FWHM, but also on their exposure depths. With this metric, the optimal convolution direction can be determined {\it a priori}, depending only on the FWHM and depth of the images. This not only simplifies the process, but also makes it more robust and less prone to creating sub-optimal difference images due to the wrong choice of the convolution direction. As an additional benefit, for a large set of images, we define a Figure-of-Merit based on this metric, which allows us to rank a list of images and determine the ones best suited to be used as templates, thus streamlining and automating the data reduction process.

\end{abstract}


\section{Introduction} \label{sec:intro}

In recent decades, wide-field time-domain surveys such as PS1 \citep{Chambers2016}, ZTF \citep{Bellm2019A}, and ATLAS \citep{Tonry2018} have dramatically increased the observation rate of optical transients. These surveys use difference image analysis \citep[DIA;][]{Phillips1995} to discover transients and extract their light curves. The basic idea of DIA is to subtract one image from another, thus removing all static astrophysical objects like nonvariable stars, galaxies, and nebulosity from the image. Any significant deviation from zero in these difference images is either astrophysical variability or contamination by instrumental and/or data processing artifacts. The success of DIA strongly depends on the ability to match one image to the other before subtraction, in particular, to match the point spread functions (PSFs) which may vary both spatially and with time.

\citet{Phillips1995} and \citet{Tomaney1996} first introduced the idea of using a convolution kernel to match the PSFs from one image to another. However, their method relied on the determination of the PSFs from isolated, bright stars which proved to be slow, inefficient, and not robust. A breakthrough came with the pioneering work of \citet{Kochanski1996}, \citet{Alard1998}, and \citet{Alard2000}, who developed a framework to determine the convolution kernel without the explicit knowledge of the PSFs. This method works both in sparse extragalactic fields as well as crowded fields like the Galactic plane or globular clusters. DIA has been successfully used by many transient survey programs \citep[e.g.,][]{Cao2016,Price2019}. Several commonly used implementations of DIA are programs known as \hotpants\ \citep{Becker_hotpants}, {\tt ZOGY} \citep{Zackay2016}, and {\tt SFFT} \citep{Hu2022}.

When the PSFs of two images are matched, a decision needs to be made, which image is convolved to match the other. The general convention is to convolve the sharper image since it is easier to smooth out a PSF than the converse, the so-called deconvolution which can introduce unwanted artifacts \citep{Phillips1995}. This means that the convolution direction is either determined {\it a priori}, for example by comparing the seeing of the images, or by using statistical algorithms within the DIA implementations to determine the convolution direction automatically. The automatic selection of convolution direction used in \hotpants\ works correctly most of the time, but is still one of the dominant failure modes (i.e., producing poor image subtractions).

Moreover, we have found that the sharpness of an image is not the only relevant parameter that determines the correct convolution direction; the relative depth of the images is also relevant. In this paper, we determine how relevant the depth parameter is for DIA. There are many cases where images with a significant difference in depth produce better difference images when the deeper image is convolved to match the shallower one, even if the PSF is broader.

In this work, we present a quantitative way to determine the optimal convolution direction based on {\it a priori} information about the PSF and depth of the images. With such a quantitative method, we can make the reduction process more robust and decrease the number of false positives due to reduction artifacts. It also allows us to automate and improve upon one of the most time-consuming and cumbersome processes in DIA: template selection. Optimizing template selection is important since it impacts the quality of measurements we can make (i.e., photometry). A poor choice of template can greatly reduce the S/N of the difference images and limits what can be measured, hence impacting the science that can be accomplished. To test this optimization, we need a large data set with images covering a wide range of seeing and depth values.

We describe the data used in this study and their reduction in Sec.~\ref{sec:data}. In Sec.~\ref{sec:method}, we describe our method and apply it to example data sets. We verify and discuss our results in Sec.~\ref{sec:results}. We summarize and discuss the potential of our methods in Sec.~\ref{sec:discussion}.

\section{Observations and Data Reduction} \label{sec:data}

\subsection{Observations}

We used images taken with DECam on the Blanco 4-meter telescope located at Cerro Tololo Inter-American Observatory (CTIO) in Chile \citep{Honscheid08,Flaugher15}. DECam is a 570 megapixel camera consisting of 60, $2048 \times 4096$ pixel CCDs in a hexagonal array. The FOV of DECam is about $3~\mathrm{deg}^2$ with a pixel scale of $0.27''$. In order to probe fields with different stellar density, we selected extragalactic fields from the Young Supernova Experiment (\citep[YSE,][]{Jones2021}) DECam survey (fields 257A and 403C; PI: Rest, Program IDs: 2021A-0275 and 2023A-237157, \citealp{Rest2022}) and fields in the Galactic plane from the DECam Eta Car survey (fields ec0814 and ec0915; PI: Rest, Program IDs: 2021B-0325 and 2023A-643849). For this analysis, we use images taken in the {\it griz} filters. Details of these fields are provided in Table~\ref{tab:fields}.

\begin{table}
    \centering
    \caption{DECam Field Coordinates }
    \begin{tabular}{|c|cc|}
    \hline
    Field & RA [$^\circ$] & DEC [$^\circ$] \\
    \hline
    257A & \phantom{1}52.46 & \phantom{2}-4.20 \\
    403C & \phantom{1}85.67 & -24.29 \\
    ec0814 & 158.03 & -60.69 \\
    ec0915 & 161.26 & -59.68 \\
    \hline
    \end{tabular}
    \label{tab:fields}
\end{table}

We also apply the same analysis for data from the Swope telescope, a 1-meter telescope located at Las Campanas Observatory in Chile, in Sec.~\ref{sec:swope}. The Swope direct camera consists of one CCD of $4096 \times 4112$ pixels. The FOV is $29.7 \times 29.8~\mathrm{arcmin}^2$ with a pixel scale of $0.435''$. We chose two extragalactic fields with moderate stellar density from the Swope Supernova Survey (2022xkq and 2018jag; PI: Piro). Details of these two fields are provided in Table~\ref{tab:swope_fields}.

\begin{table}
    \centering
    \caption{Swope Field Coordinates}
    \begin{tabular}{|c|cc|}
    \hline
    Field & RA [$^\circ$] & DEC [$^\circ$] \\
    \hline
    2022xkq & \phantom{1}76.35 & -11.88 \\
    2018jag & \phantom{1}15.95 & \phantom{1}10.59 \\
    \hline
    \end{tabular}
    \label{tab:swope_fields}
\end{table}

For each field captured by DECam, there are hundreds of images available. For our analysis, we wanted to take all possible combinations of difference images for a given set of images obtained with a certain filter, but this is computationally intensive and simply not feasible for hundreds of images. Therefore we chose a subset of images, about 25 images per field for a given filter, that probe the full range of FWHM and depth. For our analysis, we selected four CCDs distributed around the DECam field rather than all 60 CCDs of DECam because the seeing and depth is expected to be similar across the focal plane for a given image. We show examples of images from fields 257A, 403C, ec0814, and ec0915 in Table~\ref{tab:images}. The full table of images will be available online with the published article.

\begin{table*}
    \centering
    \caption{Summary Data for DECam Fields\tablenotemark{a}}
    \addtolength{\tabcolsep}{-0.15em}
    \begin{tabular}{|c|rrrrrrrr|}
    \hline
    Field & Exposure ID & Observation Date & RA [$^\circ$] & DEC [$^\circ$] & Exposure Time [sec] & Filter & $\mathrm{FWHM}$ [px] & $m_{5\sigma}$ [mag] \\
    \hline
    257A & 1034043 & \phantom{} 2021-09-17 T07:27:25 & 86.35 & -24.37 & 50.0 & i & 4.01 & 22.53 \\
    257A & 1034044 & 2021-09-17 T07:28:44 & 86.35 & -24.37 & 50.0 & z & 4.39 & 22.00 \\
    257A & 1035185 & 2021-09-20 T07:34:07 & 86.35 & -24.37 & 50.0 & z & 3.70 & 22.04 \\
    257A & 1035186 & 2021-09-20 T07:35:25 & 86.35 & -24.37 & 50.0 & i & 4.85 & 22.07 \\
    257A & 1035197 & 2021-09-20 T07:51:02 & 86.35 & -24.37 & 50.0 & i & 4.10 & 22.27 \\
    403C & 1034023 & 2021-09-17 T06:59:36 & 53.08 & -4.28 & 50.0 & i & 5.48 & 22.30 \\
    403C & 1034024 & 2021-09-17 T07:00:56 & 53.08 & -4.28 & 50.0 & z & 4.53 & 22.06 \\
    403C & 1034437 & 2021-09-18 T08:11:53 & 53.08 & -4.28 & 50.0 & i & 3.70 & 22.73 \\
    403C & 1034438 & 2021-09-18 T08:13:11 & 53.08 & -4.28 & 50.0 & z & 4.00 & 22.27 \\
    403C & 1035149 & 2021-09-20 T06:54:27 & 53.08 & -4.28 & 50.0 & i & 4.34 & 22.12 \\
    ec0814 & 178896 & 2013-02-18 T07:28:44 & 158.06 & -60.76 & 200.0 & g & 5.39 & 23.65 \\
    ec0814 & 179490 & 2013-02-20 T05:26:55 & 158.05 & -60.77 & 100.0 & i & 3.99 & 23.22 \\
    ec0814 & 184940 & 2013-03-09 T05:22:11 & 158.04 & -60.77 & 100.0 & i & 3.59 & 23.39 \\
    ec0814 & 219195 & 2013-07-15 T23:26:51 & 158.03 & -60.77 & 100.0 & r & 3.70 & 23.26 \\
    ec0814 & 272821 & 2014-01-10 T04:53:42 & 158.04 & -60.77 & 100.0 & i & 3.23 & 23.02 \\
    ec0915 & 155041 & 2012-11-24 T07:24:47 & 161.32 & -59.75 & 100.0 & r & 3.88 & 23.10 \\
    ec0915 & 178843 & 2013-02-18 T05:03:60 & 161.24 & -59.75 & 100.0 & i & 3.94 & 23.03 \\
    ec0915 & 178883 & 2013-02-18 T06:48:55 & 161.26 & -59.77 & 200.0 & g & 5.76 & 23.12 \\
    ec0915 & 184934 & 2013-03-09 T05:06:17 & 161.28 & -59.76 & 100.0 & i & 3.47 & 23.34 \\
    ec0915 & 203991 & 2013-05-04 T23:04:31 & 161.26 & -59.77 & 100.0 & i & 4.91 & 22.42 \\
    \hline
    \end{tabular}
    \tablenotetext{a}{The full table of DECam images will be available online with the published article.}
    \label{tab:images}
\end{table*}

The analysis of Swope data also uses images in the {\it griz} filters. We show example images of these Swope fields in Table~\ref{tab:swope_images}. The full table of images will be available online with the published article.

\begin{table*}
    \centering
    \caption{Summary Data for Swope Fields\tablenotemark{a}}
    \addtolength{\tabcolsep}{-0.15em}
    \begin{tabular}{|c|rrrrrrrr|}
    \hline
    Field & Exposure ID & Observation Date & RA [$^\circ$] & DEC [$^\circ$] & Exposure Time [sec] & Filter & $\mathrm{FWHM}$ [px] & $m_{5\sigma}$ [mag] \\
    \hline
    2022xkq & 2105 & 2023-03-08 T04:24:01 & 76.48 & -11.77 & 45.0 & g & 2.69 & 21.70 \\
    2022xkq & 3110 & 2023-03-08 T04:24:27 & 76.46 & -11.75 & 45.0 & g & 2.73 & 21.79 \\
    2022xkq & 4103 & 2023-03-08 T04:24:51 & 76.47 & -11.75 & 45.0 & g & 2.89 & 21.76 \\
    2022xkq & 5094 & 2023-03-08 T04:25:16 & 76.46 & -11.77 & 45.0 & g & 3.59 & 21.58 \\
    2022xkq & 6097 & 2023-03-08 T04:25:43 & 76.46 & -11.75 & 45.0 & g & 4.98 & 21.14 \\
    2018jag & 6058 & 2023-03-12 T03:36:11 & 16.07 & 10.71 & 1200.0 & g & 3.16 & 23.60 \\
    2018jag & 6056 & 2023-03-12 T03:36:31 & 16.07 & 10.71 & 1200.0 & r & 2.83 & 23.13 \\
    2018jag & 6057 & 2023-03-12 T03:36:51 & 16.07 & 10.71 & 1200.0 & i & 2.88 & 22.44 \\
    2018jag & 3063 & 2023-03-15 T19:11:41 & 16.06 & 10.70 & 90.0 & g & 2.75 & 21.49 \\
    2018jag & 4076 & 2023-03-15 T19:12:06 & 16.05 & 10.69 & 180.0 & g & 3.25 & 21.73 \\
    \hline
    \end{tabular}
    \tablenotetext{a}{The full table of Swope images will be available online with the published article.}
    \label{tab:swope_images}
\end{table*}

\subsection{Data Reduction}

All images were reduced with the {\tt photpipe} difference image pipeline \citep{Rest2005A,Rest2014}. The input to this pipeline is the standard reduced and WCS aligned images downloaded from the NOIRlab Astro Data Archive\footnote{https://noirlab.edu/public/projects/astrodataarchive/}. {\tt Photpipe} then deprojects the images to a common tangential plane using {\tt SWarp} \citep{Bertin2002}, and performs photometry using a modified version of {\tt DoPhot}  \citep{Schechter1993}. We use the FWHM determined by {\tt DoPhot} for the following analysis. 

{\tt DoPhot} uses a truncated 2-dimensional Gaussian as the PSF model, and fits it to the objects found in the image. Each object is categorized as a star, galaxy, double star, or other. {\tt DoPhot} makes iterative passes through the image with thresholds of increasing depth, in order to find the fainter objects. The typical stellar PSF is calculated by taking the average FWHM along the major and minor axes of objects categorized as stellar. The average FWHM is accurate to first order to quantify the PSF size, even for ellipsoidal PSF shapes,  as the average FWHM will get larger for any extended axis.

We used reference catalogs to determine the zero points for the conversion of the instrumental magnitudes into apparent AB magnitudes. For the YSE DECam fields 257A and 403C, we use the PS1 catalog as the reference catalog, while for ec0814 and ec0915, we use the DECAPS survey \citep{Schlafly18}. For the Swope fields, we use PS1 catalog as the reference catalog.

We then characterize the exposure depth as $m_{5\sigma}$, assuming background-noise-dominated photometry within the {\tt DoPhot} aperture. $m_{5\sigma}$ is defined as the magnitude of point sources for which the effective counts are at S/N of $5$. Thus an image with greater depth will have a larger (fainter) $m_{5\sigma}$ value.  We note that the value of $m_{5\sigma}$ is affected both by exposure time and photometric conditions at the time of the observation. As the last step, the kernel matching and difference imaging is performed with \hotpants\ \citep{Becker_hotpants}.

\subsubsection{Difference Image Quality Metric} \label{sec:chi2}

The core part of our analysis is to compare the quality of two difference images, which are derived from a single pair of images (in the same filter) convolved in both directions. It is essential for our analysis to have a metric that robustly evaluates the quality of the difference images. One of the most straightforward ways is the normalized $\chi^2$ of the flux of a difference image itself. For a perfect difference image without any astrophysical variability nor any instrument or reduction artifacts, all flux values are consistent with zero within their errors. Thus we expect a $\chi^2=1$ for such a perfect image. In practice, the $\chi^2$ may be increased by the following sources:
\begin{enumerate}
    \item {\bf True astrophysical variability:} Astrophysical variability is in general rare, especially in extra-galactic fields, and therefore contributes only minimally to an increase in the $\chi^2$. The exceptions are dense stellar fields with many variable stars, or fields with large light echoes. In these exceptional cases, true astrophysical variability can indeed increase the $\chi^2$ significantly for both directions of convolution. The proper motion and parallax of celestial objects can also cause some offsets that may be larger depending on the time of separation between the images, which will increase the $\chi^2$ for both directions of convolution.
    \item {\bf Non-astrophysical transients or moving objects:} Objects like satellites, solar system objects, or cosmic rays all add non-subtracted flux to the difference images. In most cases, this also adds only insignificantly to the $\chi^2$. The images undergo astrometric calibration to line up the pixel to real sky locations. In general, the WCS correction is accurate to better than 0.2 arcsec. Codes like \hotpants\ can take out small astrometric errors. Large astrometric differences will increase the $\chi^2$ for both directions of convolution. Some images may have artifacts (e.g., cosmic rays and bad pixels) that get masked out in the reduction stages. If the image with the masking gets convolved, the masking will get bigger in relation to the kernel size. This means that an artifact can have a different size of masking in the difference image depending on the convolution direction. However, the presence of artifacts is random in any image, hence the overall impact to the $\chi^2$ for either direction of convolution is expected to be similar.
    \item {\bf Imperfect kernels:} Even the best PSF-kernel model cannot account for all effects. Here are a few examples:
    \begin{itemize}
        \item There is always structure in the PSF that cannot be modeled by a finite number of Gaussians, which means that for the brightest non-saturated stars, there can be residuals from unsubtracted flux. 
        \item Differences in atmospheric dispersion due to variations in color cause relative spatial shifts in astrophysical objects. This is the Differential Chromatic Refraction (DCR) phenomenon where there is a wavelength dependence on the refractive index. This also causes the occurrence of both negative and positive residuals (so-called dipoles). \citet{Alcock1999A} showed that corrections can be applied to account for DCR effects. However, this requires that the colors of the astrophysical objects are known a priori, which is very difficult to obtain, in particular for large data sets. Therefore none of the current wide-field time-domain surveys account for DCR.
        \item Very bright stars, often highly saturated, have significant spikes and halos. The structure, brightness, and orientation of these spikes and halos depend on the rotation of the telescope as well as the relative position in the focal plane. The kernel size is often significantly smaller than these effects, and therefore does not account for them. 
    \end{itemize}
    \item {\bf Wrong convolution direction:} If the wrong convolution direction is chosen, then every static astrophysical object with significant flux will have some residual flux in the difference image, e.g., the well-known deconvolution rings. These residuals are caused by the numerical instability found in deconvolution at higher frequencies where noise dominates \citep[e.g.,][]{Phillips1995,Alard1998}.
\end{enumerate}

For most images, the first three sources mentioned above increase the $\chi^2$ only by a small amount and in both directions of convolution. In addition, for images from the same region of the sky or even the same field, the effect of these three sources is most likely the same for all images. In that case, the $\chi^2$ is a good metric for the quality of the difference image.

However, for images from different astrophysical environments (e.g., sparse versus dense fields), or under different circumstances (e.g., bright satellite streak), the $\chi^2$ might be significantly larger than unity even though it is the best difference image possible. We therefore introduce $\Delta \chi^2$ as the difference between the $\chi^2$ values of both convolution directions for a given image pair. The three first sources in the above list have exactly the same impact on the $\chi^2$ of both convolution directions, so the only difference between the $\chi^2$ is due to the convolution direction. This means that $\Delta \chi^2$ is the better metric for evaluating the impact of the convolution direction on the difference image quality. To clarify, $\chi^2$ is used as the metric for the quality of the difference image and $\Delta \chi^2$ is used as a relative metric to see which convolution direction does better.

\section{Method and Analysis} \label{sec:method}

\subsection{Convolution Direction} \label{sec:conv_direct}

One of the main parameters determining the quality of difference images is the correct choice of the convolution direction. The general rule is that the image with the better PSF, i.e., the smaller PSF, is convolved to match the image with the worse PSF. However, in practice this has proven to not always be true, and implementations of DIA like \hotpants\ have often the option to determine the preferred convolution direction empirically. This empirical determination is one of the main failure modes of DIA, and our goal is to fix this failure mode using a priori information about the images. We find that besides the PSF size, the depth of the images is the other parameter that determines the preferred convolution direction. As stated before, we use the FWHM of an image to describe the PSF size, and $m_{5\sigma}$ to describe the depth. The ellipticity of the PSF is a secondary parameter, and beyond the scope of this paper.

In order to determine how the difference image quality depends on the convolution direction, we perform difference imaging in both directions on all possible pairs of images in our data set. For a given subtraction of two images, $I_{k}$ and $I_{l}$, we define the difference image $D_{kl} =  I_{k} - I_{l} \otimes K$, where $K$ is the convolution kernel, such that the image that is subtracted, $I_l$, is the one that is convolved.

\begin{figure*}
    \includegraphics[width=0.5\textwidth]{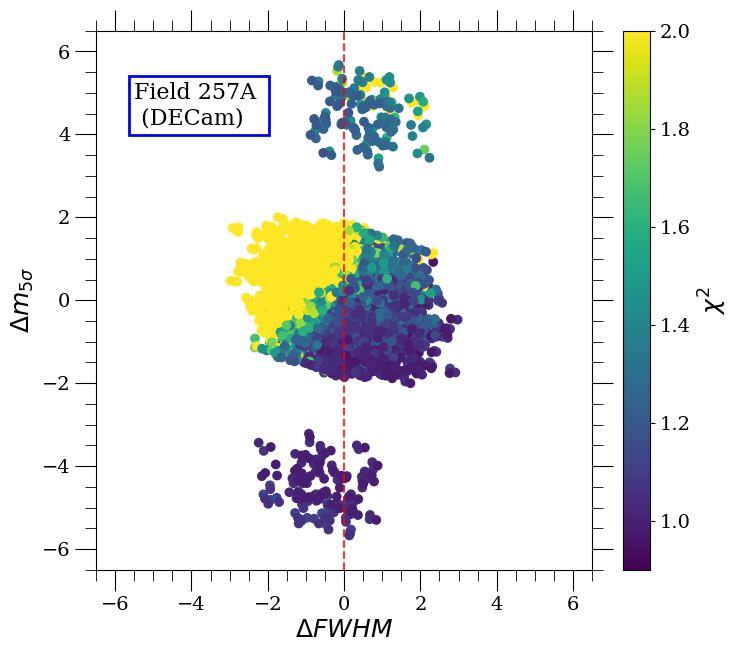}
    \includegraphics[width=0.5\textwidth]{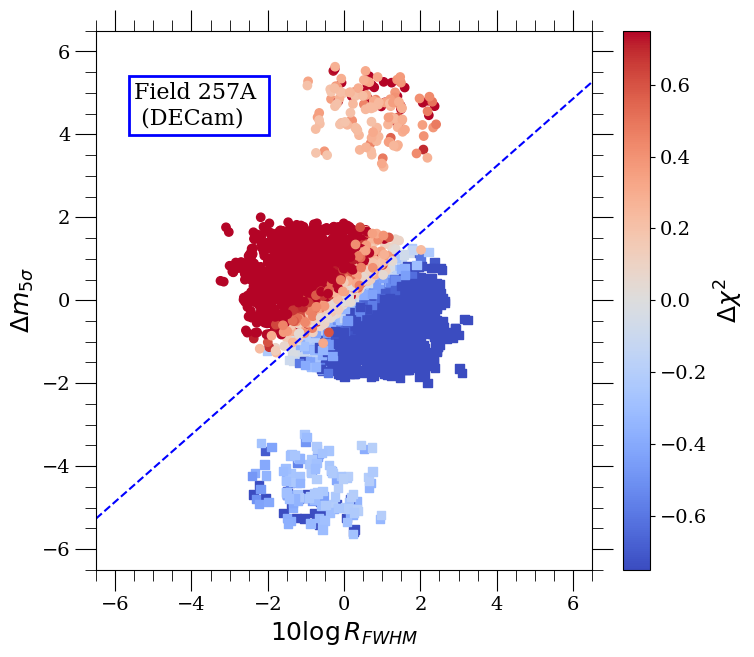}
    \caption{\textbf{Left}: Plot of $\Delta m_{5\sigma}$ vs. $\Delta${FWHM} for YSE field 257A. The color bar is the reduced $\chi^2$ of a given difference image. The standard practice would lead to the expectation that a vertical line (red dashed line) would best split the difference image pairs.  However, this is clearly not the case. \textbf{Right}: Similar plot for the same field, but with revised axes: $\Delta m_{5\sigma}$ vs. $10 \log R_{\mathrm{FWHM}}$. These plots may look similar, but the relation of FWHM is changed to a log ratio and the color bar is now the $\Delta \chi^2$ with a different color map, which better shows which difference image had the better convolution. The difference images with a better convolution direction or, more accurately, have a lower $\chi^2$ are represented by square markers on the plot. The blue dashed line is the best fit line to divide the better difference images from their counterparts for field 257A.}
    \label{fig:257A_standard}
\end{figure*}

We investigated the relative values of the seeing and depth for all pairs of images and their resulting quality of difference images. In the left panel of Fig.~\ref{fig:257A_standard}, we first use the YSE Field 257A to demonstrate our method. We compute $\Delta \mathrm{FWHM} = \mathrm{FWHM}_k - \mathrm{FWHM}_l$ ($x$-axis) and $\Delta m_{5\sigma} = m_{5\sigma,k} - m_{5\sigma,l}$ ($y$-axis) for each unique image pair and convolution direction with the corresponding values of $\chi^2$ indicated by the colorbar. Under the standard assumption that the better-seeing image should be convolved to match the worse image \citep[e.g.,][]{Phillips1995}, we would expect to see a clear delineation in the $\chi^2$ values as a function of only $\Delta \mathrm{FWHM}$, i.e., a vertical line at $\Delta \mathrm{FWHM} = 0$ would cleanly divide the points with lower $\chi^2$ values on the right and the higher values for the corresponding mirror difference images (with the convolution direction flipped) on the left. We can readily see for Field 257A that this is not the case. Instead, we find that a line of finite slope is needed to cleanly separate the preferred convolution direction for each image pair from its mirrored counterpart.

In the right panel of Fig.~\ref{fig:257A_standard}, we refine this approach in a few ways to make it more generally applicable. First, we use $\Delta \chi^2$, i.e., the difference in $\chi^2$ for a given image pair with the convolution run in both directions, as our quality metric (e.g., color bar). As described above in Sec.~\ref{sec:chi2}, this removes any contributions to $\chi^2$ between image pairs and across different fields that are not related to the choice of convolution direction. We also replace $\Delta \mathrm{FWHM}$ with a dimensionless ratio to remove the dependence on units and the pixel scale of the imager. Lastly, we take the logarithm and multiply by a factor of 10 so that this quantity, $10 \log R_{\mathrm{FWHM}} = 10 \log_{10} (\mathrm{FWHM}_k/\mathrm{FWHM}_l)$, matches the same scale as the $\Delta m_{5\sigma}$ parameter. This is the parameterization we will use going forward.

In addition to using $\chi^2$ as a quality metric for difference images, we verified the quality of subtraction by investigating how well stars were subtracted in the difference images. Using the star catalog, the pipeline can perform ``forced'' photometry in the difference images at positions where each star was subtracted out. Determining the flux and its uncertainty (noise) at each position of a star subtraction can reveal the extent of any residual flux, thus measuring the quality of the subtraction. The uncertainty of the flux is referred to as the dflux in this analysis. Assuming Gaussian stellar profiles, the stars should be subtracted out such that the distribution of flux over its uncertainty (dflux) at the subtracted stars' positions is a Gaussian distribution centered at $0$ with a standard deviation ($\sigma_{flux/dflux}$) of $1$ due to Poisson statistics. The multitude of faint stars typically subtract out such that any residuals are at or below the noise level. Brighter stars are more likely to leave observable residuals due to imperfections in PSF matching. Thus, the standard deviation of the distribution is expected to increase above unity for brighter stars. A relatively poor difference image will have more significant residuals and show a strong increase in $\sigma_{flux/dflux}$ for brighter stars, whereas a good difference image will show only a modest rise in $\sigma_{flux/dflux}$.

\begin{figure}
    \centering
    \includegraphics[width=0.5\textwidth]{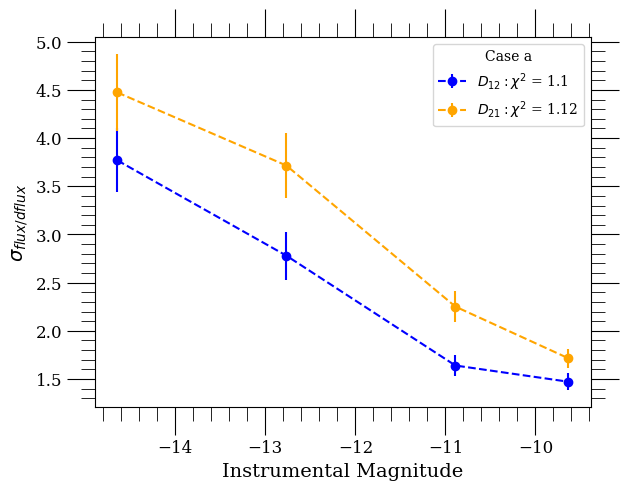}
    \includegraphics[width=0.5\textwidth]{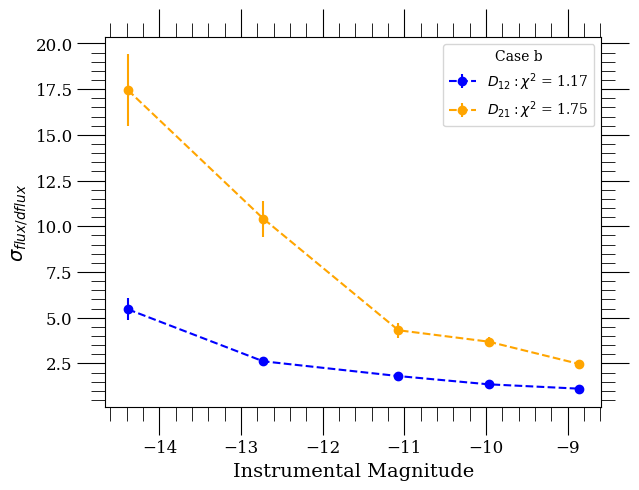}
    \caption{Plots of $\sigma_{flux/dflux}$ per magnitude bin measured in the difference images at locations of subtracted stars for example cases $a$ (top) and $b$ (bottom) from YSE field 257A; see Table~\ref{tab:257A_sel} and Fig.~\ref{fig:257A_sel}. Case $a$ (top plot) is for the case with a pair of images with comparable seeing and depth. Case $b$ (bottom plot) is for the case with a pair of images with varying seeing and depth. Clearly, when the $\chi^2$ of two difference images are similar then the $\sigma_{flux/dflux}$ trend is also similar. When $\chi^2$ varies, the $\sigma_{flux}$ trend varies as well, with the residual noise much lower in the image with lower $\chi^2$.}
    \label{fig:257A_stats}
\end{figure}

Fig.~\ref{fig:257A_stats} demonstrates the expected behavior of $\sigma_{flux/dflux}$ per instrumental magnitude bin for a couple of examples of difference images. These examples are for cases of different quality of input images that are described in Sec.~\ref{sec:results}; see Table~\ref{tab:257A_sel} and Fig.~\ref{fig:257A_sel}. Case $a$ has similar quality of input images and case $b$ is of varying seeing and depth input images. The difference image with the worse $\chi^2$ is expected to have a steeper increase in $\sigma_{flux/dflux}$ as the instrumental magnitude becomes brighter (more negative). For case $a$, the difference images have similar $\chi^2$ values and show similar $\sigma_{flux/dflux}$ distributions. In contrast, for case $b$, the difference images have different $\chi^2$ values and show large differences in the $\sigma_{flux/dflux}$ distribution, with the lower $\chi^2$ difference image clearly indicating the better subtraction. In these examples, the $\sigma_{flux/dflux}$ distribution correlates well with $\chi^2$, verifying our use of $\chi^2$ as a quality metric for difference images. We are more confident that $\Delta \chi^2$ correctly emphasizes the relative differences between opposing convolution directions of difference images.

\begin{figure*}
    \centering
    \includegraphics[width=0.45\textwidth]{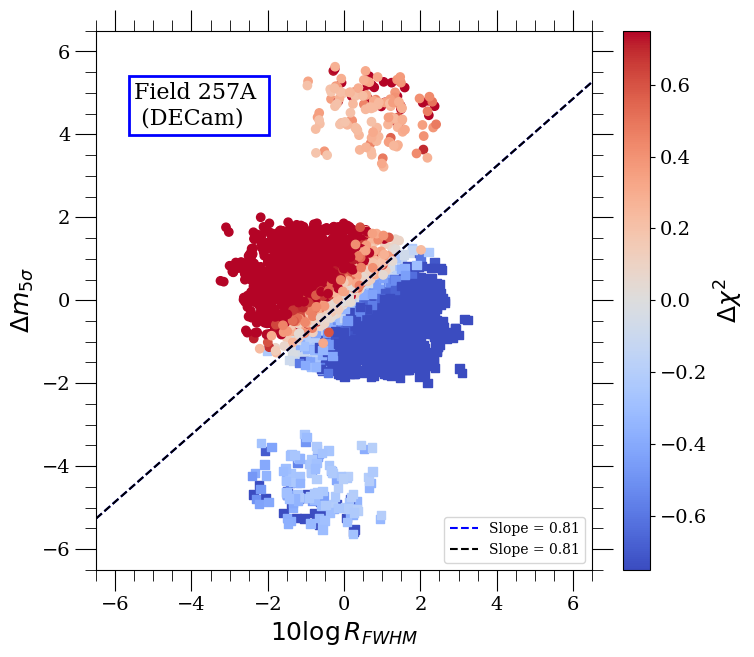}
    \includegraphics[width=0.45\textwidth]{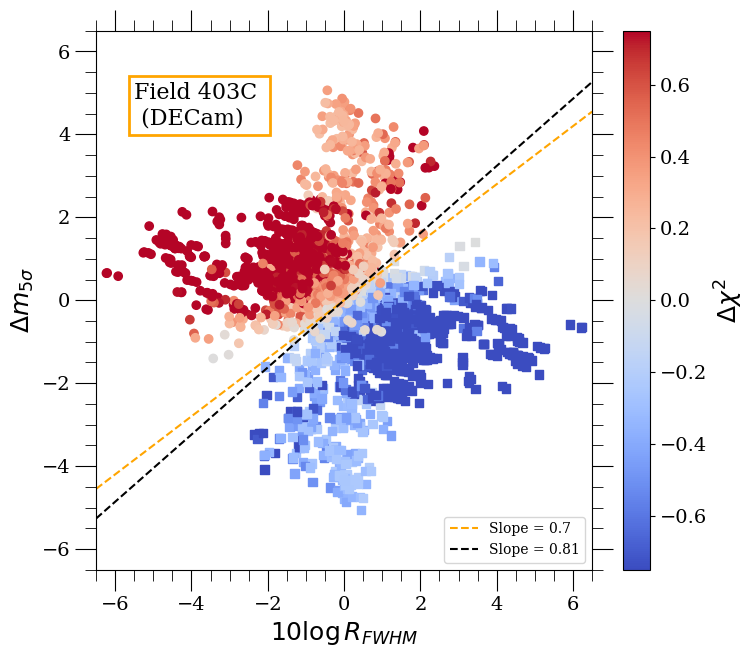}
    \includegraphics[width=0.45\textwidth]{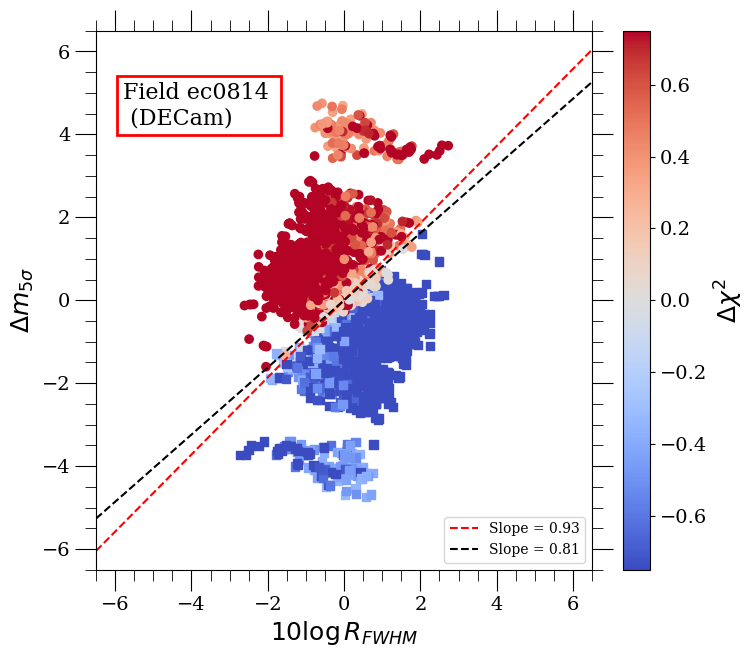}
    \includegraphics[width=0.45\textwidth]{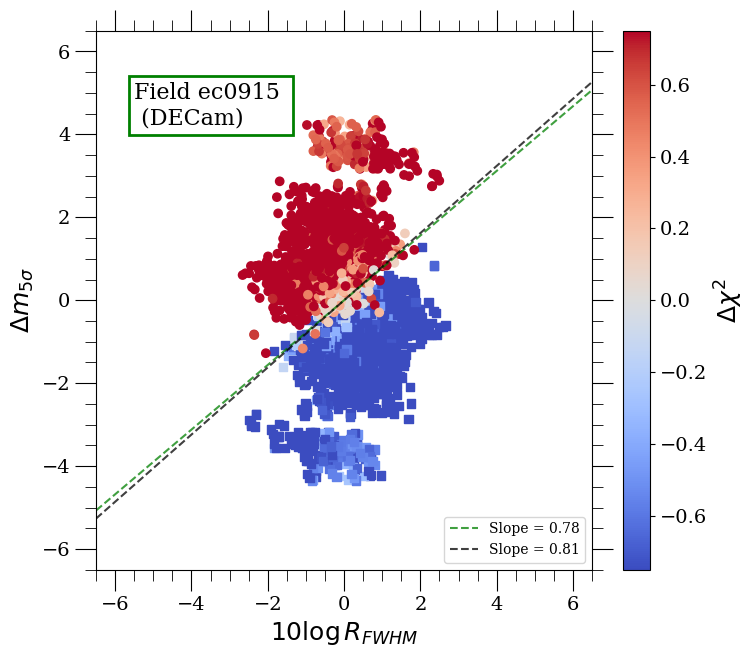}
    \includegraphics[width=0.45\textwidth]{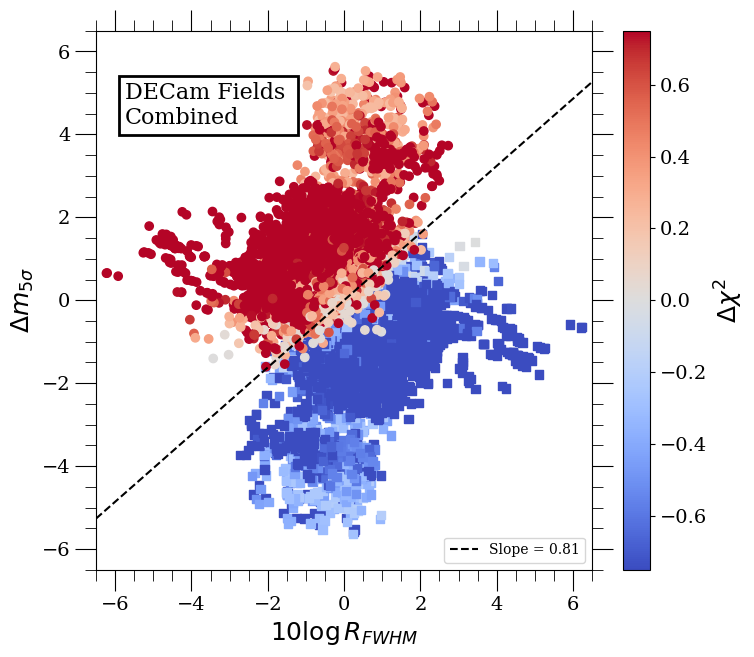}
    \caption{Plots similar to the right panel of Fig.~\ref{fig:257A_standard} for all of the DECam fields, including a plot combining all fields. The colored lines are the respective fields' best fit lines that divide the better convolutions from their worse counterparts. The black line represents the global best fit line when combining all DECam fields; see Table~\ref{tab:results} for the slope values. These plots strongly show that a sloped line, not a vertical line, best splits the convolution direction. Hence, depth is a significant parameter to consider when it comes to image convolution.}
    \label{fig:m5vsfwhm_all}
\end{figure*}

As the criteria for the analysis, we applied two conditions for the data we considered. At least one direction of convolution for a given pair of images should result with a difference image with $\chi^2 < 2.0$, to ensure we don't include images that will always fail. The other condition is that we only include data where the difference images' $\chi^2$ value correlates correctly with the their $\sigma_{flux/dflux}$ behavior. Meaning, the difference image with a lower $\chi^2$ should also have a shallower increase in its $\sigma_{flux/dflux}$ distribution. Difference images may have contradictory $\chi^2$ values and $\sigma_{flux/dflux}$ distributions if the $\chi^2$ is impacted more by empty sky then the locations of stars. A good difference image will have both good background and star subtraction, so we exclude difference images that fail one or the other. The rest of the paper continues with the subset of data that satisfies these conditions. In Fig.~\ref{fig:m5vsfwhm_all} we show the $\Delta m_{5\sigma}$ vs. $10 \log R_{\mathrm{FWHM}}$ plots for the other fields in our data set. All fields agree that a sloped line better divides the convolution direction choice.

We determined the best fit line that separates the convolution direction by first defining a `contamination' parameter. Contamination is the percentage of good difference images compared to total images found on one side of the line on a $\Delta m_{5\sigma}$ vs. $10 \log R_{\mathrm{FWHM}}$ plot. Specifically, contamination is calculated as the count sum of good images divided by the count sum of all images found on one side of the dividing line. The sums are weighted by the $\Delta \chi^2$ values of each difference image.

\begin{figure}
    \includegraphics[width=0.5\textwidth]{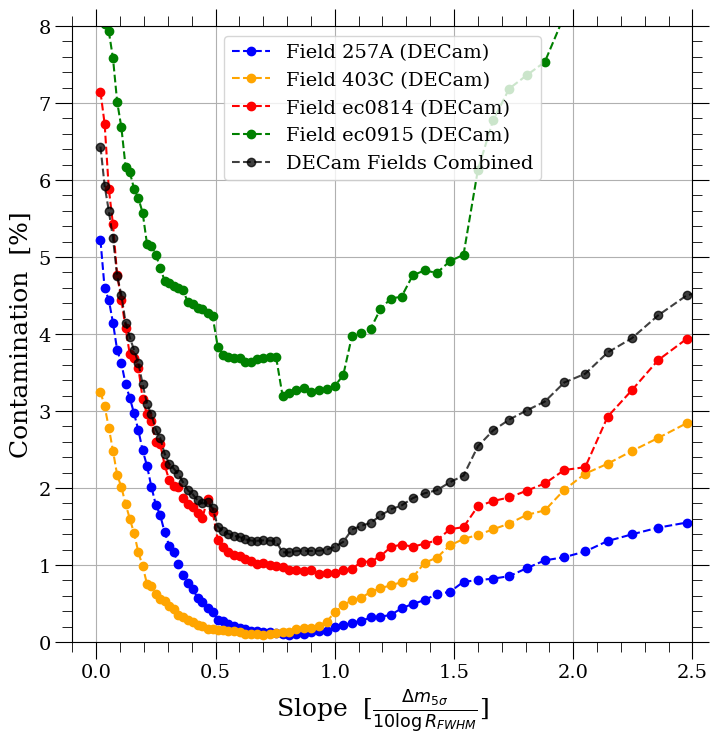}
    \caption{Plot of the contamination for lines of varying slope for each field and the combination of all the DECam fields. The x-axis is the slope of the line and the y-axis is the corresponding contamination percentage (see text). The lines are of form $y=mx$ on a $\Delta m_{5\sigma}$ vs. $10 \log R_{\mathrm{FWHM}}$ plot, where the slope is in form of $\frac{\Delta m_{5\sigma}}{10 \log R_{\mathrm{FWHM}}}$ with units of mags. The minimum of contamination varies slightly per field, but has considerable overlap between $0.35 - 1.47$. It is clear a sloped line better divides the plot, hence bolstering the claim that depth of an image has a significant impact on the better direction of convolution.}
    \label{fig:combined_slope_dist}
\end{figure}

In reality, most pairs close to the dividing line are close in $\chi^2$ (i.e., $\Delta \chi^2$ is small), but we still want to choose the lower value result when possible.  By varying the slope of the line and determining the contamination, we found the best slope line as well as a range of slopes where the contamination is lowest per field. Ultimately by combining the data of all the DECam fields, we found the slope value that can be applied globally for DECam images. We plot the contamination for each line over varying slopes in Fig.~\ref{fig:combined_slope_dist}. We expect a similar minimum even with the different type of fields.

In Table~\ref{tab:results} we list the slope that gives the minimum contamination for each field, as well as the respective range of slopes that keep contamination within $1\%$ of their minimum. 
We include the results for combining all DECam fields. We find considerable overlap between the fields with a slope range of $0.35 - 1.47$.

\begin{table}
    \centering
    \caption{The slopes for the best dividing lines on the $\Delta m_{5\sigma}$ vs. $10 \log R_{\mathrm{FWHM}}$ plot for each DECam field.}
    \begin{tabular}{|c|cc|}
    \hline
    Field & Range & Best fit \\
    \hline
    257A & 0.33 - 1.96 & 0.81 \\
    403C & 0.18 - 1.44 & 0.70 \\
    ec0814 & 0.44 - 1.58 & 0.93 \\
    ec0915 & 0.49 - 1.11 & 0.78 \\
    Combined & 0.35 - 1.47 & 0.81 \\
    \hline
    \end{tabular}
    \label{tab:results}
\end{table}

\subsection{Analysis of the Swope Data} \label{sec:swope}

The analysis above concentrated on the DECam data.  Here, we
follow a similar analysis for images taken with the Swope telescope to demonstrate how the technique applies to other data sets as well. In addition to the criteria described in Sec.~\ref{sec:conv_direct}, we excluded difference images with bad offsets due to imperfect stitching of the amplifiers in these data sets. We continue the analysis of Swope with these new subsets.

\begin{figure}
    \centering
    \includegraphics[width=0.45\textwidth]{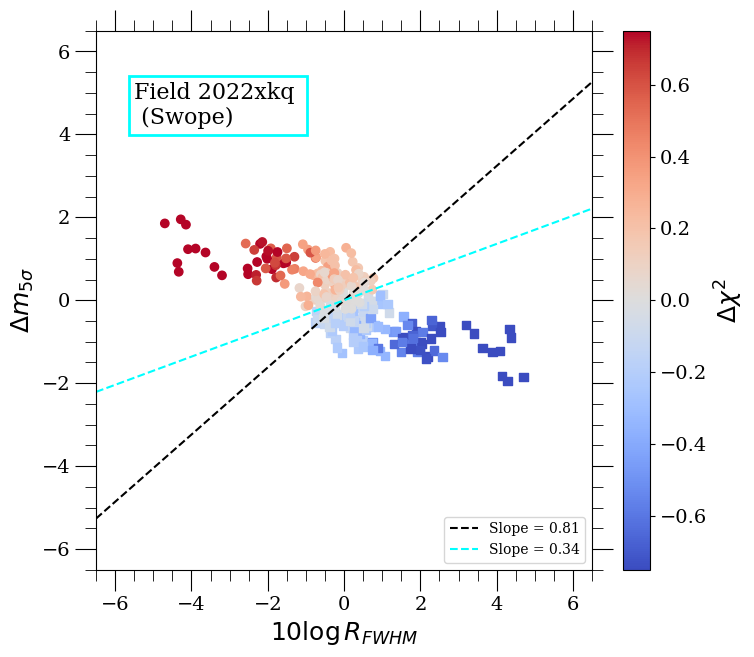}
    \includegraphics[width=0.45\textwidth]{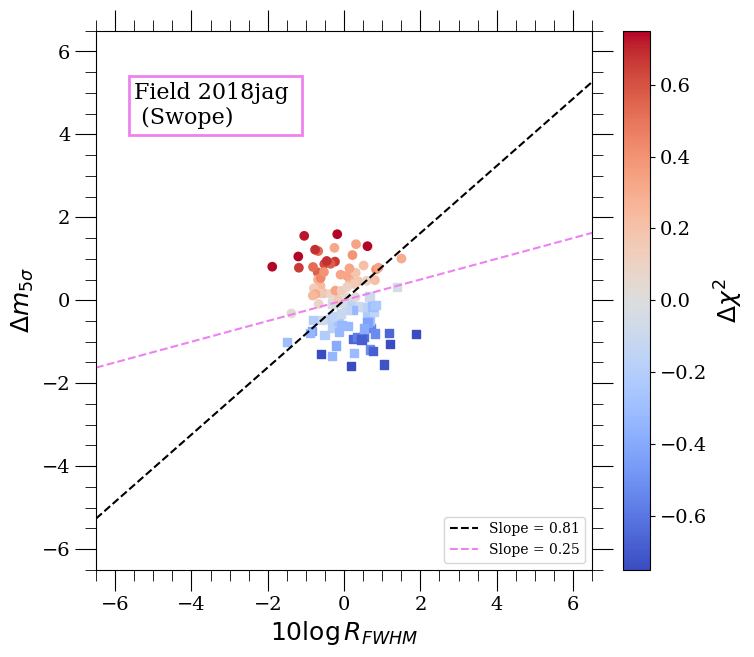}
    \caption{Plots similar to the right panel of Fig.~\ref{fig:257A_standard} for the Swope fields. The colored lines are the respective fields' best fit lines that divide the better convolutions from their worse counterparts. The black line represents the global best fit from the DECam fields. Despite the different best fit slopes, it is clear that a sloped line, not a vertical line, best splits the convolution direction. Hence, image depth is a significant parameter to consider when it comes to image convolution in this case as well.}
    \label{fig:swope_m5vsfwhm}
\end{figure}

We show the $\Delta m_{5\sigma}$ vs. $10 \log R_{\mathrm{FWHM}}$ plots for the Swope fields in Fig.~\ref{fig:swope_m5vsfwhm}. The Swope fields show that a significantly shallower slope best splits the plot, implying that the depth difference of the input images is a more significant factor than seeing. We determined the best fit lines for the Swope fields similarly to the procedure used for the DECam fields.

We see that the best fit lines for the Swope fields have a slope close to $0.3$ whereas the best fit line for the DECam fields have a slope close to $0.8$. Applying the same slope of $0.8$ to Swope fields would still mostly get the correct convolution direction and especially out perform the previous method of only using seeing to decide the direction. We acknowledge that one will need to do their own analysis if they want to optimize the slope for a certain instrument. However, regardless of instrument, it is clear that image depth has a significant impact on the determination of the best convolution direction.

\subsection{Template Selection}

To automate the template selection, we need to define a Figure-of-Merit (FoM) that quantifies each image's quality as the template for a group of images. We define the FoM as the perpendicular distance between the data point on a $\Delta m_{5\sigma}$ vs. $10 \log R_{\mathrm{FWHM}}$ plot and the dividing line. We follow the formula for the distance between a point and a line \citep[e.g.,][]{Math_Formulas} to define the FoM, defined by the following equation:

\begin{equation}  \label{eq:fom}
    FoM = \frac{\Delta m_{5\sigma,r} - m_s*10log_{10}(R_{FWHM,r})}{\sqrt{1+m_s^2}}
\end{equation}

Here $m_s$ is the slope of the dividing line. We choose a slope of $0.81$ as the global best fit slope found in Fig.~\ref{fig:combined_slope_dist} for all DECam fields. The $\Delta m_{5\sigma,r}$ and $R_{FWHM,r}$ denotes the reference point of depth and seeing. We use a reference point based on the median seeing and depth for a given field, since this is in general a good representative for a given group of images. We remove the absolute sign typically found in the distance formula because we want the FoM value to become negative when the data point falls below the dividing line where we find the better difference images. Hence, the image with the smaller FoM is better suited to be the template for any given pair of images. This process produces a ranked list of potential template images from which the best template can be selected for a given situation. Note that the highest ranked template will generally produce a good difference image for any given image of the group, but there can be other images in the ranked list that will produce a more suitable difference image, depending on the science application; see in Sec.~\ref{sec:discussion}. In many cases, these secondary changes are modest and the highest ranked template can just be applied directly. We have publicly available code to employ this method\footnote{\url{https://github.com/AnguloRodrigo/DIA_TemplateSelection}}.

\section{Results} \label{sec:results}

\begin{figure}
    \includegraphics[width=0.5\textwidth]{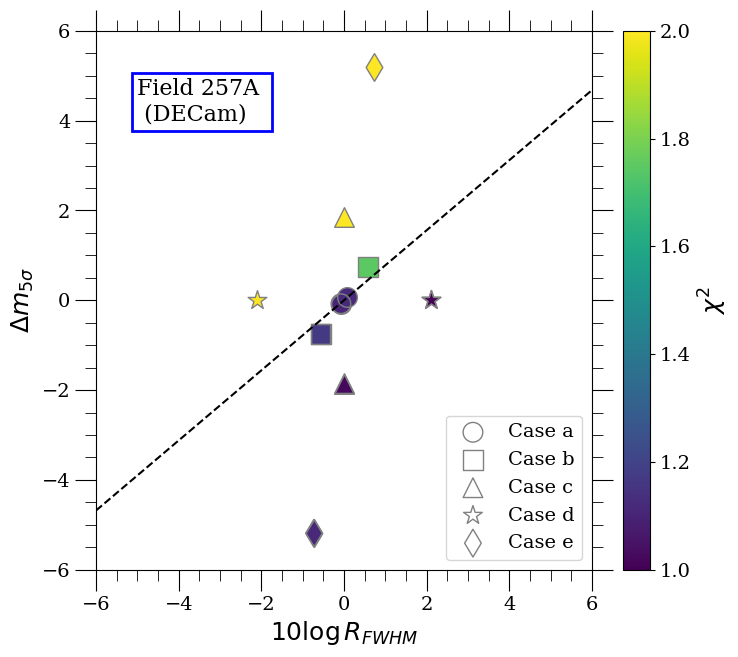}
    \caption{Plot for YSE field 257A as in right panel in Fig.~\ref{fig:257A_standard}, but now excluding all data except for the example cases discussed in the text; see Table~\ref{tab:257A_sel}. The example cases sample differing regions of the $\Delta m_{5\sigma}$ vs. $10 \log R_{\mathrm{FWHM}}$ plot.}
    \label{fig:257A_sel}
\end{figure}

We now take a closer look into some of the difference images of YSE field 257A. We choose some examples that cover various image quality combinations: ($a$) images with comparable seeing and depth, ($b$) images with varying seeing and depth, ($c$) images with only varying depth, ($d$) images with only varying seeing, and ($e$) images with large difference in seeing and depth. See Table~\ref{tab:257A_sel} for the relevant values of these example cases. Fig.~\ref{fig:257A_sel} shows where these cases fall on the $\Delta m_{5\sigma}$ vs. $10 \log R_{\mathrm{FWHM}}$ plot and more clearly shows the mirroring of each corresponding data point. 

\begin{table}
    \centering
    \caption{Examples from YSE Field 257A}
    \begin{tabular}{|c|cccccc|}
    \hline
    Case & $\chi^2_{12}$ & $\chi^2_{21}$ & $\mathrm{FWHM}_1$ & $\mathrm{FWHM}_2$ & $m_{5\sigma,1}$ & $m_{5\sigma,2}$ \\
    \hline
    $a$ & 1.10 & 1.12 & 4.97 & 5.06 & 22.33 & 22.41 \\
    $b$ & 1.17 & 1.75 & 4.11 & 4.68 & 22.56 & 23.30 \\
    $c$ & 1.02 & 2.33 & 3.57 & 3.57 & 22.04 & 23.90 \\
    $d$ & 0.95 & 3.49 & 4.76 & 2.92 & 22.94 & 22.94 \\
    $e$ & 1.12 & 2.17 & 2.91 & 3.44 & 17.61 & 22.80 \\
    \hline
    \end{tabular}
    \label{tab:257A_sel}
\end{table}

We inspected the actual difference images for the cases shown in Fig.~\ref{fig:257A_sel}. For each difference image, we can inspect how well stars are subtracted out and the uniformity of the background. For each case, we show the input images before convolution, referred as the science images, and the two possible difference images. The grayscale limits of the images are $5$ times the noise level of the image above and below the background sky value. The input images are scaled to the same photometric zero-point, however the limits are different due to different night conditions (e.g., amount of moonlight present) when the images were taken. For display purposes, the difference images are scaled to the same limits. We denote our images such that $D_{12}$ corresponds to the preferred convolution direction according to our method.

\begin{figure*}
	\includegraphics[width=1.0\textwidth]{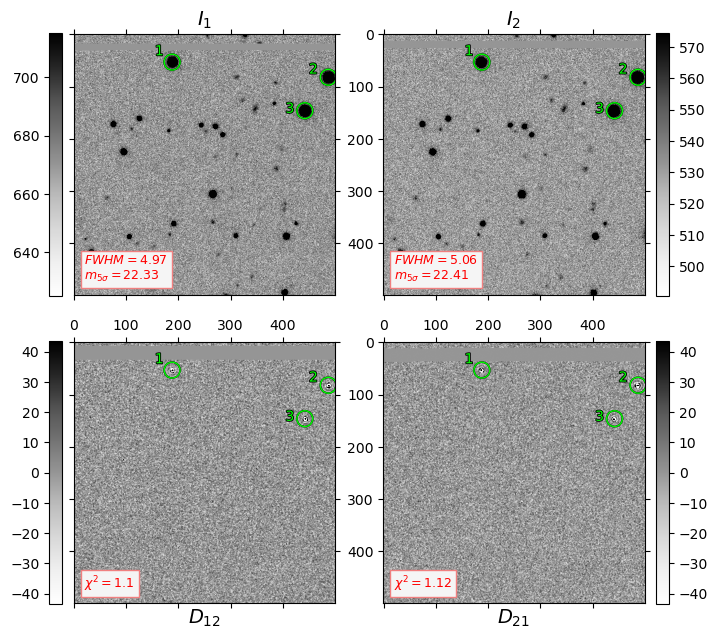}
    \caption{Images for example case $a$ in YSE field 257A. The top two images are the science images (before convolution) and the bottom two images are the difference images of both convolutions. The grayscale limits of the images are $5$ times the noise level of the image above and below the background sky value. The science images are scaled to the same photometric zero-point, but have different limits due to different night conditions (e.g., amount of moonlight present). For display purposes, the difference images are scaled to the same limits. These images are all cropped to the same location where one can see how well the stars were subtracted; some objects are labeled with a circle to help with the comparison. $D_{12}$ has $I_2$ as the template and $D_{21}$ has $I_1$ as the template. These images are of comparable seeing and depth, so as expected, both produce similar results. Some residuals are left behind by the star subtraction, shown in the circles, but the effect is small seen by the low $\chi^2$ values.}
    \label{fig:257A_a}
\end{figure*}

As a ``sanity'' check, we check case $a$ where a pair of images with comparable seeing and depth give similar difference images for either direction of convolution. We show the science images and difference images in Fig.~\ref{fig:257A_a}. As expected, the choice of convolution direction is of little importance for near identical images. We see that both difference images have good star subtraction, have a uniform background, and have low $\chi^2$ values. In this case, $D_{12}$ does have a lower $\chi^2$ than its counterpart, albeit marginally so. The numbered circles emphasize locations to easily compare the subtractions. Both difference images have some residuals left over from the star subtraction, but is marginal when compared to the images as a whole.

\begin{figure*}
	\includegraphics[width=1.0\textwidth]{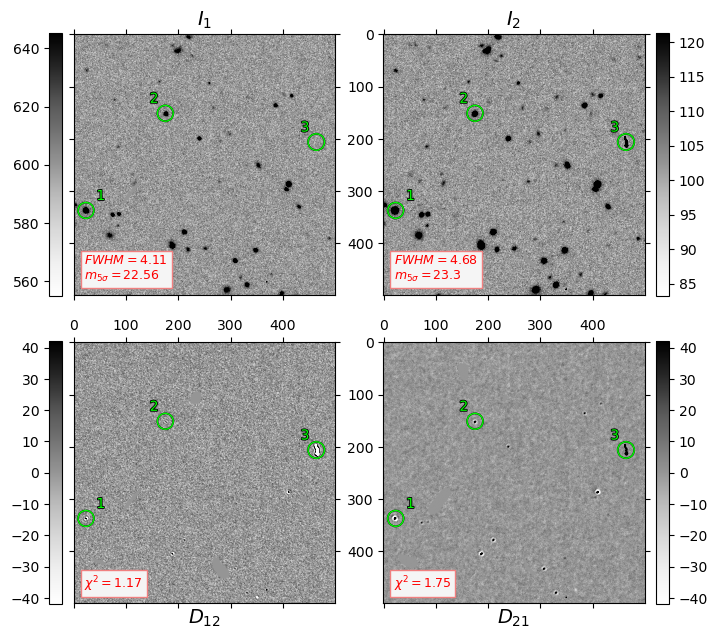}
    \caption{Images for example case $b$ in YSE field 257A. Same description as Fig.~\ref{fig:257A_a}. This is a case where the images have different seeing and depth. $D_{21}$ is the result of convolving a better seeing image to the worse seeing image in the standard way. $D_{12}$ is the convolution of a worse seeing image, but with better depth to a better seeing image. Clearly $D_{12}$ is the better difference image with overall less residuals, see circle 1 and 2. Circle 3 follows an artifact that only shows up in image 2 and hence appears in both difference images.}
    \label{fig:257A_b}
\end{figure*}

The next case is an example of how convolving a better seeing image may actually result in the worse difference image. Case $b$ has a pair of images with both varying seeing and depth. We show the science images and difference images in Fig.~\ref{fig:257A_b}. Here, our method of convolving a worse seeing, but better depth image ($D_{12}$) produces the better difference image. The numbered circles follow points of interest to compare the difference images. $D_{21}$ clearly has the worse residuals as seen in circles 1 and 2, and has the worse $\chi^2$ value. The circle 3 is included to show how artifacts show up in both difference images. This case is a good example in support of our method.

\begin{figure*}
	\includegraphics[width=1.0\textwidth]{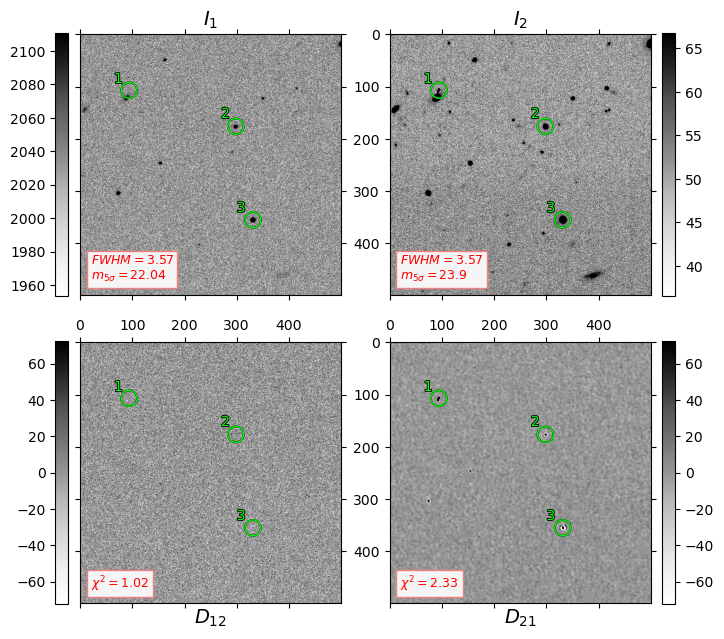}
    \caption{Images for example case $c$ in YSE field 257A. Same description as Fig.~\ref{fig:257A_a}. Here the images have comparable seeing, but varying depths. We see that $D_{21}$ has worse star subtraction that leaves significant residuals, shown in circles 2 and 3. Circle 1 follows an artifact that shows up in $I_2$, but not $I_1$. We expect to see this artifact in both difference images, but the artifact is masked in $D_{12}$. $D_{12}$ is the better difference image with good star subtraction and a uniform background.}
    \label{fig:257A_c}
\end{figure*}

To examine the significance of image depth, we investigated case $c$ where a pair of images with comparable seeing but varying depth are convolved. We show the science images and difference images in Fig.~\ref{fig:257A_c}. Here, we see that convolving the deeper image ($D_{12}$) produces a cleaner subtraction with a lower $\chi^2$. $D_{21}$, in particular, has significant residuals from subtraction artifacts related to the poor removal of stars shown in circles labeled 2 and 3. $D_{12}$ has also successfully masked an artifact that was present in $I_2$ and not in $I_1$, whereas the artifact is still present in $D_{21}$; see circle 1. $D_{12}$ is the better difference image to search for the true astrophysical changes.

\begin{figure*}
	\includegraphics[width=1.0\textwidth]{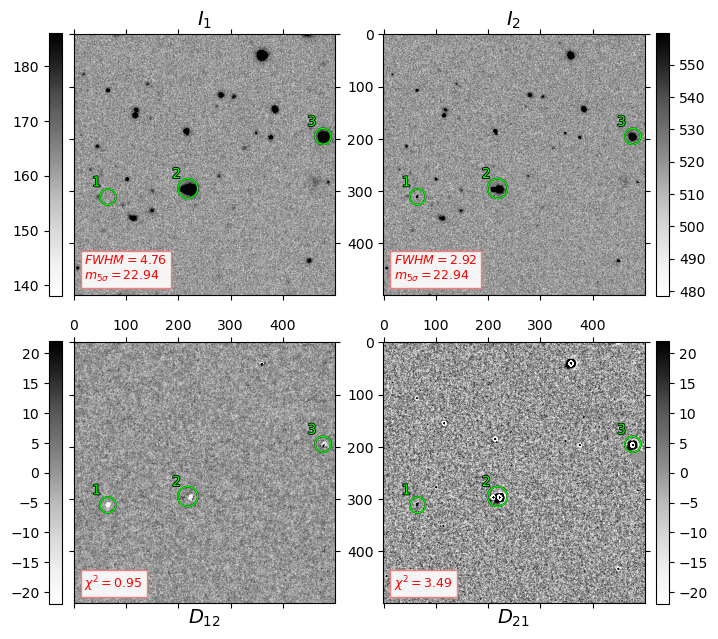}
    \caption{Images for example case $d$ in YSE field 257A. Same description as Fig.~\ref{fig:257A_a}. The images have comparable depths, but varying seeing. Here our method and the standard practice agree since both would choose $I_2$ (better seeing image) as the template. $D_{21}$ is the worse difference image because it has bad star subtraction with PSF-ringing and a noisy background; see circles 2 and 3. $D_{12}$ shows some residuals in circles 2 and 3 likely due to bright stars and the significant seeing difference, but still clearly the better difference image. Note, circle 1 follows an artifact that only shows up in image 2 and hence appears in both difference images.}
    \label{fig:257A_d}
\end{figure*}

We next check the case where the images differ only in seeing. Case $d$ has a pair of images with comparable depth, but varying seeing. We show the science images and difference images in Fig.~\ref{fig:257A_d}. Here, our method agrees with the standard practice to choose the image with better seeing as the template when the pair of images have similar depths. $D_{21}$ clearly has worse star removal with prominent ring-like residuals around the badly-subtracted stars; see circles labeled 2 and 3 in Fig.~\ref{fig:257A_c}. These are the well-known deconvolution rings and are the main reason why standard practice avoids convolving a worse seeing image to match a better seeing image. Here, we see that $D_{12}$ is the better difference image where the residuals left over in circles 2 and 3 are not significant to the $\chi^2$ value.

\begin{figure*}
	\includegraphics[width=1.0\textwidth]{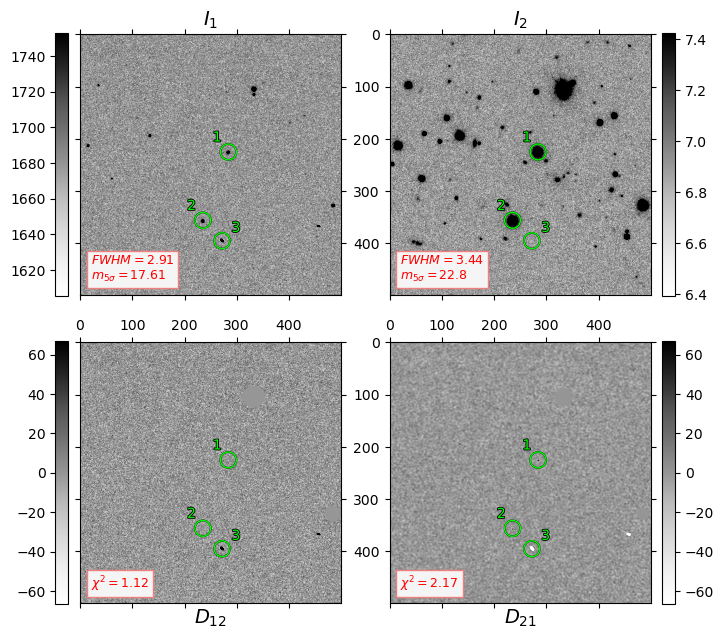}
    \caption{Images for example case $e$ in YSE field 257A. Same description as Fig.~\ref{fig:257A_a}. This is a case where the images are completely different in both seeing and depth. $D_{21}$ is the result of convolving a better seeing image to the worse seeing image in the standard way. $D_{12}$ is the convolution of a worse seeing image to a better seeing image, but with significantly better depth. It is not as clear from a glance which difference image is better, but $D_{21}$ does leave behind residuals in circles $1$ and $2$ whereas $D_{12}$ does not. Circle 3 follows an artifact that only showed up in $I_1$ and so appears in both difference images.}
    \label{fig:257A_e}
\end{figure*}

Lastly, we investigate case $e$ where the pair of images have large differences in both depth and seeing. We show the science images and difference images in Fig.~\ref{fig:257A_e}. This case also supports our method as the difference image that convolves a worse seeing, but better depth image ($D_{12}$) results with a cleaner subtraction and a better $\chi^2$ value. In Fig.~\ref{fig:257A_e}, we see that in circles $1$ and $2$ there are more residuals in $D_{21}$ than in $D_{12}$. Circle $3$ showcases how an artifact that only showed up in $I_1$ will be left over in both difference images.

These example cases confirm that our method consistently selects the better convolution direction and hence, the better difference image. Applying this methodology thus provides a way to automate template selection when bulk processing many images. We can create a ranked list based on the FoM for each field per CCD and per filter, indicating from best to worse the image's quality as a template. We aim to use this method to be able to reduce a large dataset in bulk and look for any variability that can be tied to real astrophysical phenomena. Hence, our criteria for template selection is to choose the image that is best suited to be the template for most images of a given field. Generally, defining the FoM with respect to the median values of seeing and depth of a group of images simplifies the calculations and returns accurate template choices. In certain cases, a different combination of images may produce a ``better" difference image; however, they may only be nominally better and/or the template for said difference image may not be as good a choice of template for the rest of the data set. In the next section, we discuss some specific science cases where additional template criteria need to be accounted for.

\section{Discussion and Conclusions} \label{sec:discussion}

With massive imaging data sets being generated by many current and future wide-field time-domain surveys, efficient and automated processes for reducing these data need to be applied. All these surveys use some variation of difference image analysis (DIA) to remove all static objects so that it is easier to identify and measure transients and other variable objects. One of the core principles of DIA is that the PSF of one image is matched to another before the subtraction. With kernel-matching code like \hotpants\ \citep{Becker_hotpants} and {\tt SFFT} \citep{Hu2022}, this means that a decision needs to be made of which image is matched to the other using a convolution kernel, the so-called convolution direction. The convolution direction can be determined empirically; for example, \hotpants\ has an option to determine the convolution direction by statistically comparing preliminary results in both directions. This empirical method does not work perfectly and can be one of the main failure modes when applied to large data sets.

Another way to determine the convolution direction is to use \emph{a priori} knowledge of the PSF size of the images. A general rule that historically has been applied is that the image with the smaller PSF is the one that should be convolved. However, we find that the image depth is an equally important factor in determining the convolution direction. Utilizing the normalized $\chi^2$ as the goodness of fit parameter, we find that there is a clear linear relation that can be used to determine which image is the one to be convolved to produce the best difference image. With this relation, the convolution direction can be automatically selected solely based on \emph{a priori} information, which removes one of the main failure modes of DIA using kernel matching.

The recently developed DIA algorithm {\tt ZOGY} does not use kernel matching, but instead applies statistical methods to create a so-called proper difference image whose pixel noise is uncorrelated \citep{Zackay2016}. The main assumption is that the images are background-noise dominated, and that the noise is Gaussian and independent. Since {\tt ZOGY} does not use the conventional kernel matching DIA technique, but instead relies on good prior knowledge of the PSFs of both the science and template images, it has the advantage of not needing to find the preferred direction, i.e., no decision of a convolution direction needs to be made. However our method removes this disadvantage for conventional kernel matching DIA techniques by determining the preferred direction {\it a priori}. 

We have quantified this process of template selection using not just the stellar FWHM values, but also an indicator of exposure depth. As shown in Fig.~\ref {fig:257A_standard}, a sloped line clearly divides the data into two populations on a $\Delta m_{5\sigma}$ vs. $10 \log R_{\mathrm{FWHM}}$ plot; the better difference images and their worse counterparts. We find this behavior consistent among fields of various stellar densities ($\eta$ Car vs.\ YSE fields fro DECam) and different instruments (DECam vs. Swope). The exact slope that describes the dividing line for each field may differ somewhat given the specifics of the instrument, but the overall behavior is in agreement that depth has a significant impact on determining the best convolution direction. The DECam fields show agreement in slope between very different fields (for instance, between extragalactic fields and fields along galactic plane where the stellar density vastly differs). DECam fields produce slopes $\frac{\Delta m_{5\sigma}}{10 \log R_{\mathrm{FWHM}}}$ between $0.35 - 1.47$ while the Swope fields have slopes between $0.25 - 0.35$. The two Swope fields have similar minima, but the slope is shallower than that of the DECam fields. This may be due to instrumental differences such as how the PSF is sampled and the different pixel scale. However, we can see in Fig.~\ref{fig:swope_m5vsfwhm} that applying the same slope we found from the DECam fields to Swope fields will still give the majority of the correct convolution directions. We are confident that applying the slope we found for the DECam fields to other instruments and data sets will give a significant improvement in automating the convolution direction compared to the standard practice. However, if one is truly interested in deriving an  optimal slope, then one should do an analysis similar to what we have shown above for the specific instrument.

We have developed a quantitative Figure of Merit (FoM) based on FWHM and exposure depth that can be used to determine the preferred convolution direction for a pair of images. With this FoM, we can create a quantitative ranked list of images in their suitability as a template. In the simplest case, the highest ranked image can be chosen as the template for difference image analysis. In other cases, additional criteria may play a role in selecting a template.
As an instructive example, for some surveys it is advantageous that a template is selected from the first observing season. With the ranked list in hand, the best template from the first season can be compared to the best templates from the rest of the survey. If the template from the first season is only marginally worse than the best template from the other seasons, then it is not a problem to use that template. However, if the FoM difference is large, then it might be better to use an image from a later season as the template. 

Another example would be for imagers with a large FOV with many detectors (such as DECam), where some possible templates from other observing programs may only partially overlap the pointing of interest. With the ranked list, multiple different templates can be chosen for different parts of the focal plane/detectors. In some cases as well, a master template is created as a mosaic or stack of input images. When creating mosaics, there needs to be a balance between depth and PSF sharpness. Our quantitative FoM quickly determines which images are of similar quality and thus are suitable for creating a master template. Normally this master template will always be the better choice to convolve in order to not change the science image too much. Application of the techniques described in this paper to stacked images may be helpful but will require additional work beyond the scope of this paper to assess fully.


\vspace{5mm}
\facilities{DECam (CTIO:4m), Swope (Las Campanas:1m)}

\software{
\texttt{Photpipe} \citep{Rest2005A,Rest2014},
\texttt{Dophot} \citep{Schechter1993},
\texttt{HOTPANTS} \citep{Becker_hotpants} \footnote{https://github.com/acbecker/hotpants},
\texttt{Swarp} \citep{Bertin2002},
\texttt{astropy} \citep{astropy2013,astropy2018,astropy2022},
\texttt{matplotlib} \citep{Matplotlib2007}, 
\texttt{SciPy} \citep{SciPy2020}, 
\texttt{NumPy} \citep{NumPy2020},
\texttt{pandas} \citep{reback2020pandas, mckinney-proc-scipy-2010}
}


\bibliography{refs}{}
\bibliographystyle{aasjournal}

\end{document}